\documentclass[intlimits,twoside,a4paper]{article}

\usepackage{amsmath,amssymb}
\usepackage{graphicx}

\usepackage[T2A]{fontenc}
\usepackage[cp1251]{inputenc}

\usepackage[eqsecnum]{cmpj2}

\issue{2014}{17}{2}{23702}
\doinumber{10.5488/CMP.17.23702}

\title[Obtaining the oscillator strength by reflectance spectra]%
{Determination of oscillator strength of confined excitons in a semiconductor microcavity}

\author[E.A. Cotta, P.M.S. Roma]{E.A. Cotta\refaddr{label1,label2},
P.M.S. Roma\refaddr{label1}}
\addresses{
\addr{label1} Departamento de F\'isica, Universidade Federal do Amazonas, Manaus, Brazil
\addr{label2} Instituto Nacional de Ci\^encia e Tecnologia em Nanodispositivos Semicondutores (INCT-DISSE), Brazil
}

\date{Received February 22, 2014}

\begin{document}

\maketitle

\begin{abstract}
We have achieved a significant experimental Rabi-splitting (3.4~meV) for confined polaritons in a planar
semiconductor $\lambda$ microcavity for only a single quantum well (SQW) of GaAs (10~nm) placed at the antinode.
The Rabi-splitting phenomena are discussed in detail based on the semiclassical theory, where two coupled harmonic
oscillators (excitons and photons) are used to describe the system. In this way, we can obtain the dispersion curve
of polaritons, the minimum value for the cavity reflectance and the oscillator strength to reach the strong coupling regime.
This approach describes an ensemble of excitons confined in a SQW and includes a dissipation component.
The results present a weak coupling regime, where an enhanced spontaneous emission takes place, and a
strong coupling regime, where Rabi-splitting in the dispersion curve can be observed. The theoretical
results are confronted with experimental data for the reflectance behavior in resonant and off-resonant
conditions and present a great accuracy. This allows us to determine the oscillator strength of the confined
excitons in the SQW with great precision.
\pacs 78.67.De, 42.25.Hz, 42.50.Ct, 42.50.Pq, 42.55.Sa, 42.70.Qs
\end{abstract}

\section{Introduction}

Ideally, a microcavity is a system in which a light-emitting material can interact with a single
cavity-resonant-mode or no interactive electromagnetic modes within the material transition width.
Thus, enhanced or suppressed spontaneous emission can be seen in this system, and in a cavity with
a very high quality factor $Q$, an even spontaneous oscillatory emission can be induced \cite{Morin,Haroche-a}.
The couplings between electronic excitations and optical modes have been of considerable interest
in atomic systems, which allows us to make a direct connection with optical cavities as well as with
optically active semiconductors in semiconductor microcavities. The emission and reflection spectra
of atoms in optical cavities are known to exhibit splitting due to the coupling of their dipole
transitions with the excitations of the vacuum-radiation field \cite{Agarwal-a,Zhu}. These coupled
modes are called vacuum-field Rabi splitting, and are one of the basics of quantum electrodynamics
in a cavity. In the case of semiconductor microcavities, a strong field-matter interaction occurs
between the optical modes and excitons, whose modes are relatively sharp, so the exciton-photon
modes are often called cavity polaritons \cite{Weisbuch,Gutbrot}. These couplings are important
for understanding such effects as optical bistability \cite{Cotta-a} and laser action \cite{Butov},
besides developing optoelectronic and photonic devices \cite{Cotta-b}.

Cavity polaritons are quasi-particles that are created due to a strong exciton-photon coupling in a
semiconductor microcavity. The cavity photon mode has a nearly parabolic in-plane dispersion and
can be described as the one having an effective mass which is usually a few times smaller than the exciton mass.
However, in the strong coupling regime, at resonance between the cavity mode and the exciton state, the
eigenmodes present an anticrossing between the exciton and confined photon modes resulting in the
appearance of two polariton branches (called Upper and Lower Polariton Branches), spectrally separated by
the Rabi splitting energy $\hbar \Omega$.

Very high-$Q$ optical micro- and nano-cavities present an extremely low photon loss rate in a
significant small cavity mode volume, offering surprisingly rich physics, spanning many areas of research
including nonlinear optics, quantum optics, and device physics \cite{Vahala,Noda}. Different microcavity
architectures presenting a lateral field confinement such as post, disk or toroidal microcavities,
have been presented over the last decade. In all of them, the main objective is to optimize the
miniaturization on a subwavelength-scale, and minimize the diffraction-limit and the loss contributions
such as surface scattering, radiation and finite cavity mirror reflectance. Moreover, the use of different
materials based on Cd$_{x}$Mn$_{1-x}$Te, ZnSe, GaN, In$_{x}$As$_{1-x}$P have been used to optimize
solid-state carrier properties such as oscillator strength, exciton binding energy, among others.
The combination of these efforts aims to produce devices capable of emitting a single mode laser at room
temperature, where the photons present the same quantum properties that have been registered at low temperatures.
The use of multiple quantum wells and quantum dots in traditional structures or in photonic crystals
permitted the generation of samples with a very high $Q$ \cite{Martinez}, a very low laser threshold and
a very high Rabi splitting \cite{Yoshie}.

In principle, the Rabi-splitting is proportional to the square root of the exciton oscillator
strength $f$ multiplied by the overlap of the exciton wave function with the electromagnetic field.
The oscillator strength is related to the probability of the transition from the crystal ground state to
the exciton state and is proportional to the dimensionality of the system. Due to the confinement of
excitons, the binding energy has a monotonous increase from the bulk value to 2D value, and reaches a
maximum when the well width is about the exciton Bohr radius \cite{Greene}. This effect leads to more
stable exciton states and to much more prominent excitonic effects in absorption and photoluminescence.
Moreover, despite the Rabi splitting increase with $f$, it decreases with the cavity length.
When polariton effects can be neglected (weak-coupling regime), the oscillator strength per unit
area can be related to the absorption coefficient integrated over the absorption peak, as measured
by an absorption experiment. However, in a strong-coupling regime, due to the changes made in the cavity
dispersion curve by polaritons, the reflectance spectrum does not make the distinction between absorption
and a change in the reflectivity-transmission balance. Thus, a new treatment is needed to determine $f$.

In this paper, we present a comprehensive theoretical and experimental study of cavity-polariton dispersion in a single quantum well
(SQW) of GaAs (100~{\AA}) confined by a low barrier potential (Al$_{0.3}$Ga$_{0.7}$As). Using the linear semiclassical theory,
we derive analytic equations for cavity-mode dispersion and cavity-polariton eigenfrequencies.
A careful analysis of transmittance spectra, using the semiclassical model to describe the weak
and strong coupling between the confined photons and excitons, produces a good agreement between
theoretical and experimental results. Thus, we show that the strong coupling regime can be observed
in a very simple sample with heavy-hole exciton transition. As a consequence, we can directly measure the excitonic oscillator strength.


\section{The sample and experimental setup}

\begin{figure}[!t]
 \begin{center}
\includegraphics[width=0.45\textwidth]{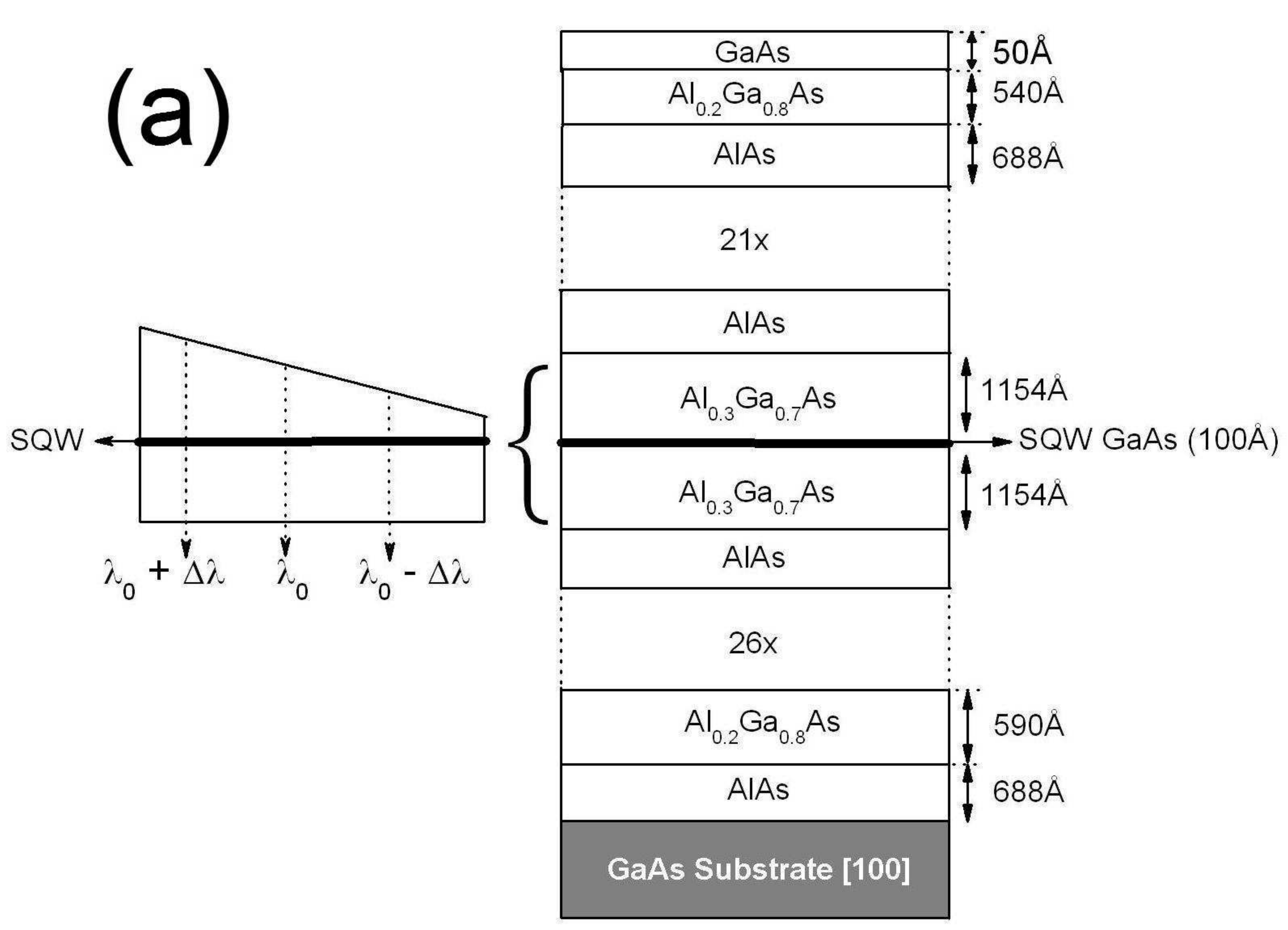} \hspace{3mm}
\includegraphics[width=0.43\textwidth]{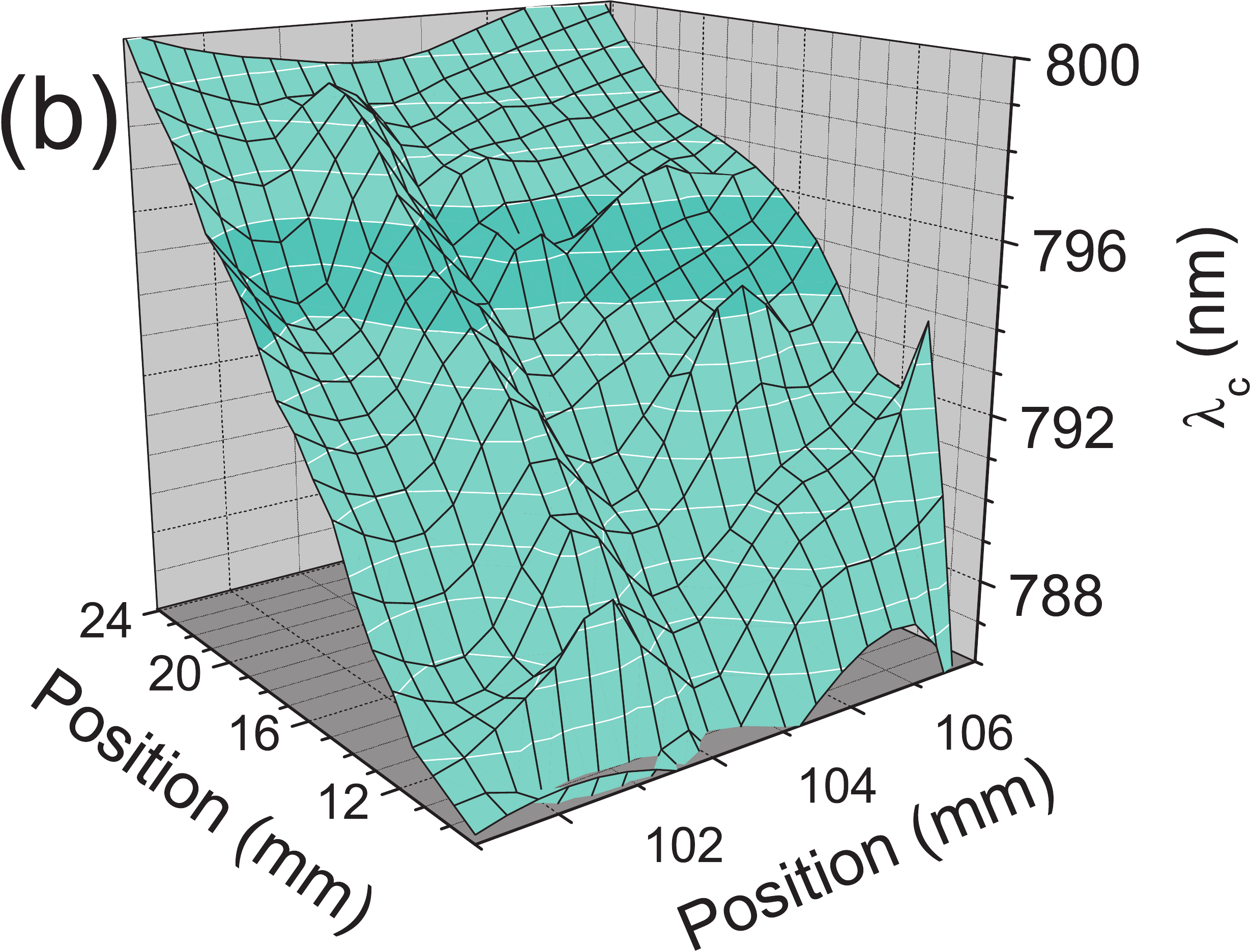} 
  \end{center}
  \caption{(a) (Color online) Scheme of the sample and (b) experimental measurement of cavity spacer layer,
  mapping the reflectance spectrum resonance in a weak-coupling regime.}
  \label{Fig:sample}
\end{figure}

In our sample, a SQW is placed at the antinode of a $\lambda$ cavity formed by two diffracted Bragg reflector (DBR)
mirrors (AlAs/Al$_{0.2}$Ga$_{0.8}$As) and kept at 10K in a cold finger cryostat. The SQW is surrounded
by Al$_{0.3}$Ga$_{0.7}$As barriers forming the spacer layer of the cavity [figure~\ref{Fig:sample}~(a)].
The sample was grown using molecular beam epitaxy technique (MBE) on GaAs[100] substrates. It rotates during
the growth of the DBR mirrors and the SQW, but stops at a specific angle for the growth of the spacer layer
on each side of the SQW, generating a thickness gradient across the sample [see figure~\ref{Fig:sample}~(b)].
This permits to perform a cavity-detuning when the sample is excited in different positions on the surface. The quality
factor of the cavity was measured, obtaining $Q \approx 3000$. The structure is designed so that the energy
of the lowest-energy heavy-hole exciton transition is the same as that of the cavity length. Since the
quantum well is strained, only the lowest-energy heavy-hole exciton is observed.

The reflectance spectrum was obtained using an unpolarized white light source, focusing it normally
on the sample surface. The measured spot size on the sample is $\sim 40$~\textmu{}m. The spatial selection
of the backscattered light is provided by imaging the sample on the entrance slit of the analysing spectrometer
with a spectral resolution of 0.5~{\AA} using a $50\%-50\%$ non-polarizing beam-splitter.

\clearpage

\section{Theoretical framework}

\subsection{Exciton-Polaritons}
Microcavity polaritons are mixed modes formed from cavity photons and dielectric excitations such as excitons.
When the microcavity contains one or several QWs, the emission from the confined electronic states (excitons)
is strongly modified by the presence of the cavity. In semiconductors, wave-vector conservation leads to the
formation of quasi-stationary exciton-polariton states, since an exciton with a given crystal momentum
interacts with one photon mode with the same wave-vector. The breakdown of crystal-momentum conservation
along the growth direction leads to an intrinsic radiative lifetime of free excitons with an in-plane
wave-vector smaller than the light wave-vector.

Due to the formation of exciton-polaritons in semiconductor microcavities, two regimes can occur:
(1) the coupling of the electromagnetic field to the crystal excitation is smaller than the
width of the cavity mode (weak-coupling regime), where irreversible decay occurs, but the emission
process is modified in relation to the free-space case; (2) the light-matter coupling is larger
than the decay rate (strong-coupling regime), the situation that can only occur for excitons, where no
irreversible decay happens. In the latter, the energy oscillates between the exciton and photon modes
(Rabi oscillations), and Rabi splitting occurs in the frequency domain. The mixed exciton-cavity modes
in the strong-coupling regime can be viewed as two-dimensional polaritons, whose radiative decay rates
are determined by the photon lifetime within the cavity.

To describe the dynamics and the radiative recombination process of confined polaritons in a SQW, we
obtain the dispersion curve by the Hamiltonian
\begin{equation}
H=\hbar\omega_{\mathrm{c},{k}} a^{\dag}_{k}a_{k}+\hbar\omega_{\mathrm{x},{k}} b^{\dag}_{k}b_{k}+\hbar\Omega\left( a^{\dag}_{k}b_{k}+b^{\dag}_{k}a_{k}\right),
\label{PolaritonHamilt}
\end{equation}
where $a_{k}^{\dag}$ and $b_{k}^{\dag}$ are creation operators for photons and excitons,
respectively, with momentum $k$. The last term in equation (\ref{PolaritonHamilt}) shows
the linear exciton-photon interaction, with the coupling energy of
\begin{equation}
\hbar\Omega = \sqrt{\frac{N_\mathrm{qw}(\hbar e)^{2}}{2\varepsilon m_0 L_\mathrm{eff}}f}\,.
\end{equation}
Here, $N_\mathrm{qw}$ is the number of QW's within the cavity, $\varepsilon$ is the permittivity
of the semiconductor ($m_0$ and $e$ is the mass and charge of the electron, respectively).
The cavity is formed by two DBR mirrors with an effective length $L_\mathrm{eff} = L_\mathrm{c}
+ \lambda_0/2n_\mathrm{c}[n_Hn_L/(n_H -n_L)]$, a refractive index $n_\mathrm{c}$ and resonant
in $\lambda_0$. The DBR mirrors consist of alternate layers, with high $n_H$ (Al$_{0.2}$Ga$_{0.8}$As)
and low $n_L$ (AlAs) refractive indexes. The additional length for the cavity in $L_\mathrm{eff}$,
comes from the energy dependence of the DBR phase, that is often much larger than $L_\mathrm{c}$, and,
therefore, produces an important Rabi-splitting reduction \cite{Savona}.

\begin{figure}[!b]
  \begin{center}
\includegraphics[width=0.48\textwidth]{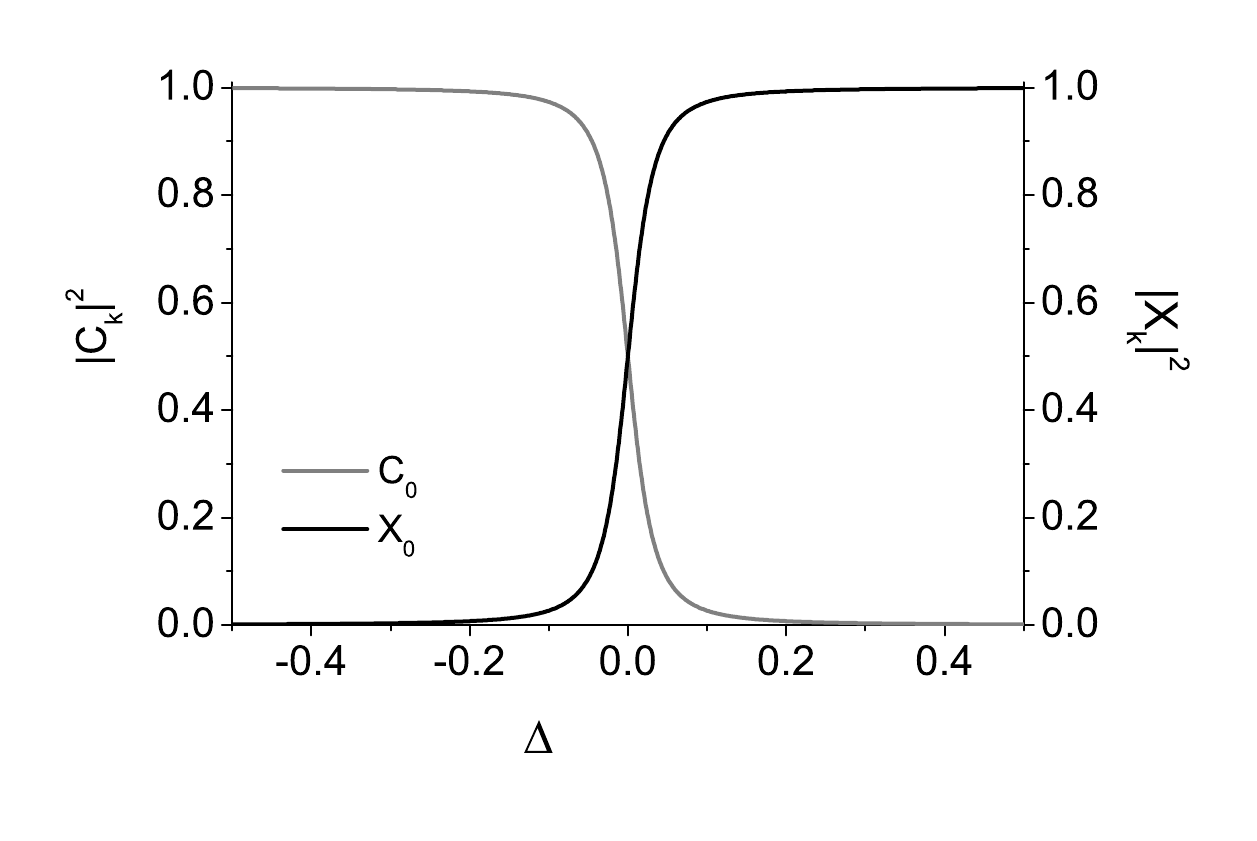}     
\includegraphics[width=0.47\textwidth]{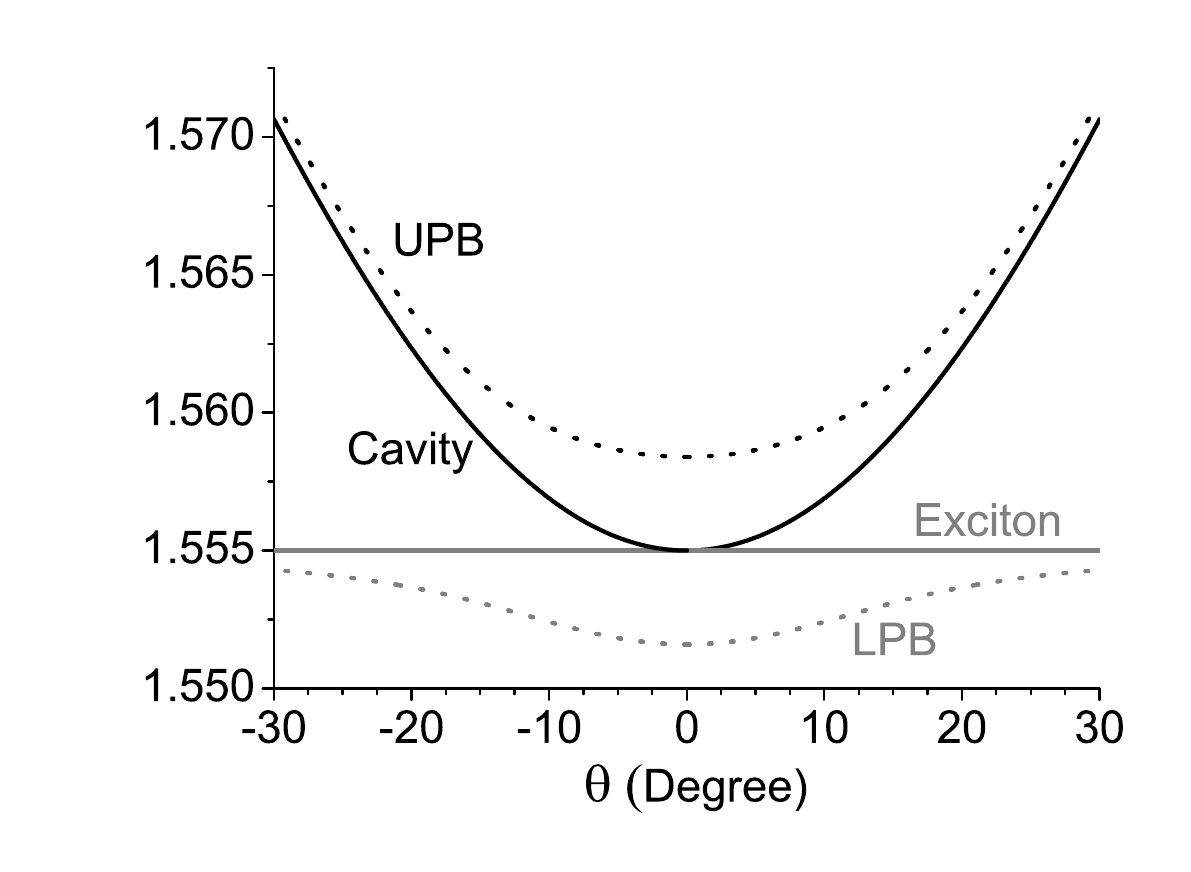}\\
\includegraphics[width=0.47\textwidth]{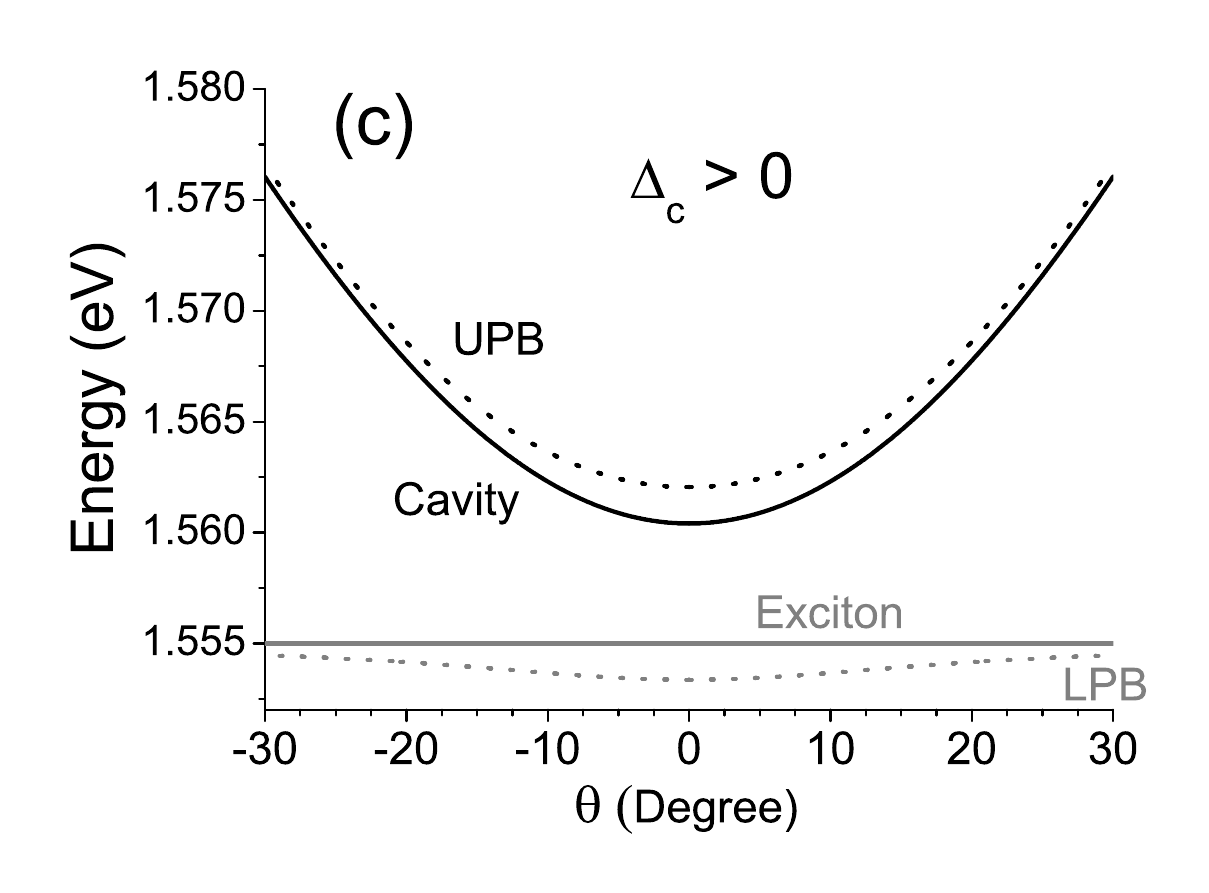}
\includegraphics[width=0.47\textwidth]{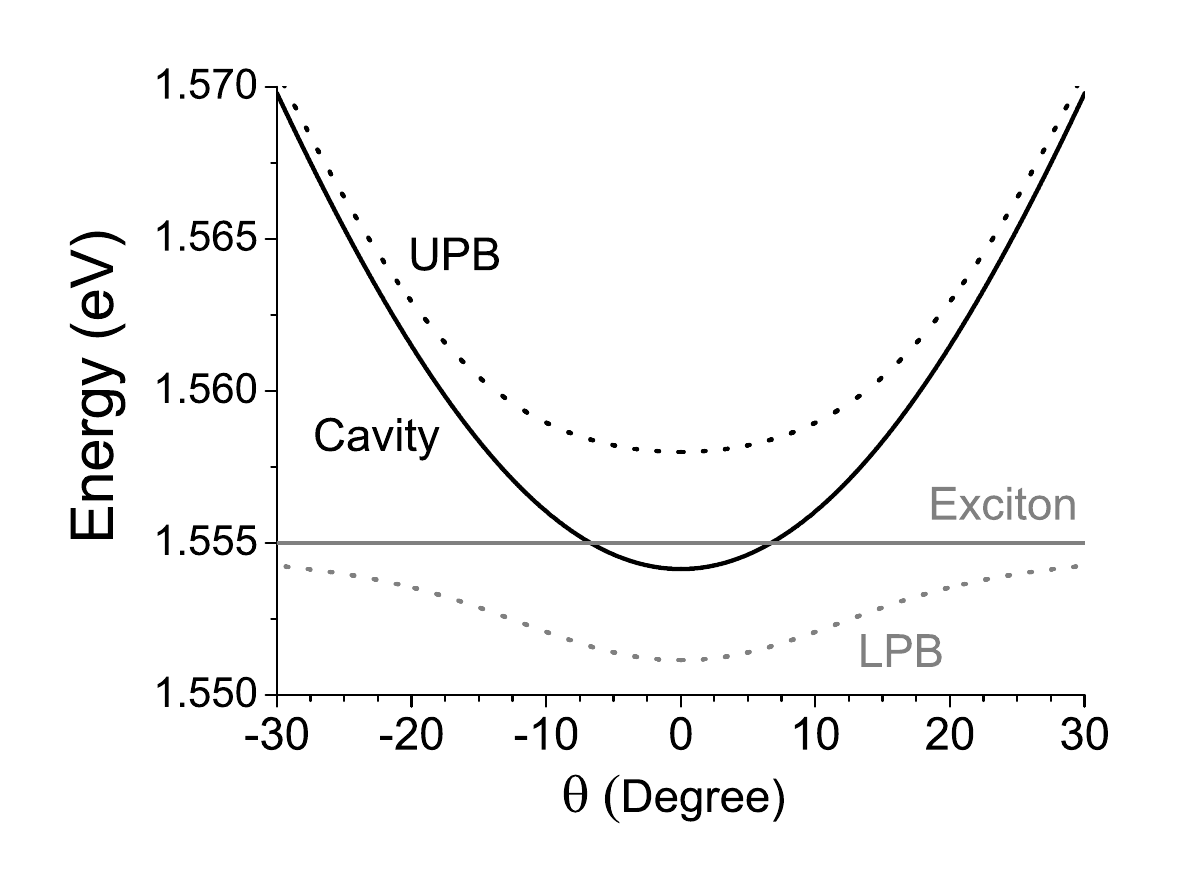}
  \end{center}
  \vspace{-5mm}
  \caption{(a) Hopfield coefficients as a function of cavity-exciton detuning
  ($\Delta_\mathrm{c}$) given by equation (\ref{HopfieldCoeff:eq}) for $\hbar \Omega = 3.4$~meV and $E_\mathrm{x} = 1.555$~eV.
  (b) Dispersion curve for confined exciton-polariton in a 100~{\AA} GaAs QW with Al$_{0.3}$Ga$_{0.7}$As barrier in resonant condition.
  (c) the same as $\Delta_{\mathrm{c},k=0}>0$, and in (d) for $\Delta_{\mathrm{c},k=0}<0$.
  The solid lines are the cavity dispersion curve (black) given by equation (\ref{Eq:CavityDisp})
  and the exciton energy (grey), that is approximately constant due to the narrow wave-vector range
  in which we analyse our data. In the strong-coupling regime, the cavity mode splits into two branches
  (dotted lines): the upper (black) and lower (grey) polariton branches.}
  \label{Fig:CavityDisp}
\end{figure}

The Hamiltonian in equation (\ref{PolaritonHamilt}) can be diagonalized using transformations
to polariton basis operators $p_k = X_k b_k + C_k a_k$ and $q_k = X_k a_k - C_k b_k$, where
\begin{eqnarray}
X_k &=& \left(\frac{\Delta_{\mathrm{c},k}+\sqrt{\Delta_{\mathrm{c},k}^2+(\hbar\Omega)^2}}{2\sqrt{\Delta_{\mathrm{c},k}^2
+(\hbar\Omega)^2}} \right) ^{1/2}, \\
C_k &=&  \left(\frac{(\hbar\Omega)^2}{2\sqrt{\Delta_{\mathrm{c},k}^2+(\hbar\Omega)^2}}\frac{1}{\left[\Delta_{\mathrm{c},k}
+\sqrt{\Delta_{\mathrm{c},k}^2+(\hbar\Omega)^2}\right]} \right) ^{1/2}
\label{HopfieldCoeff:eq}
\end{eqnarray}
are the Hopfield coefficients \cite{Hopfield}, in which $\Delta_{\mathrm{c},k}
=E_\mathrm{c}(k)-E_\mathrm{x}(k)$ is the cavity-exciton detuning. On this new basis we can verify
that an exciton-polariton is a linear superposition of an exciton and a photon state with the same
in-plane wave-number, where the probability of finding the polariton in any of these two states is
given by $|X_k|^{2}$ and $|C_k|^{2}$, that is cavity detuning dependent [see figure~\ref{Fig:CavityDisp}~(a)].

Thus, the Hamiltonian in equation (\ref{PolaritonHamilt}) can be expressed as follows:
\begin{equation}
 H = \hbar\omega_\mathrm{UP}q^{\dag}_kq_k+\hbar\omega_\mathrm{LP}p^{\dag}_kp_k\,.
\end{equation}

The interaction terms between $q_k$ (upper polariton branch~--- UPB) and $p_k$
(lower polariton branch~--- LPB) are neglected. This Hamiltonian gives us the following eigenvalues for the polariton energy
\begin{equation}
 E_{\mathrm{UP},\mathrm{LP}}=\frac{E_\mathrm{c}+E_\mathrm{x}}{2}\pm\sqrt{\hbar^2 \Omega^2+ \frac{\Delta_{\mathrm{c},k}^2}{4}}\,,
\end{equation}
where  $E_\mathrm{UP}$ and $ E_\mathrm{LP}$ are the UPB and LPB energies.
In the resonance $\Delta_{\mathrm{c},k=0} = \Delta_\mathrm{c} = 0$ [figure~\ref{Fig:CavityDisp}~(b)],
the difference of energy between LPB and UPB is given by $\hbar \Omega$. In this case, the exciton-polariton
can be generated for excitation in resonance with the cavity only at the normal angle of incidence ($\theta = 0$).
For $\Delta_\mathrm{c} > 0$ [figure~\ref{Fig:CavityDisp}~(c)] we cannot obtain a resonance condition for
any $\mathbf{k}_{||}$, and the exciton-polariton is not generated.  In the case where $\Delta_\mathrm{c} < 0$
[figure~\ref{Fig:CavityDisp}~(d)], the excitation energy must be higher than the cavity mode, out of the narrow
range of the angle of incidence ($\theta \gtrsim \pm 7^\circ$) to generate a polariton with $\mathbf{k}_{||} \neq 0$.

\subsection{Semiclassical description of collective Rabi-splitting}

The electromagnetic field stored in an optical resonator is such that the electric field $\mathbf{E}_0$
corresponding to the ground state energy of the resonator mode is extremely large \cite{Haroche-b}.
The value of $\mathbf{E}_0$ is obtained by taking the electromagnetic energy $\epsilon_0\mathbf{E}_0^{2}V$
stored in the volume $V$ of the cavity mode as equal to the energy of half a photon, $\hbar \omega/2$.
Thus, we obtain $\mathbf{E}_0 = \sqrt{\hbar \omega/(2\epsilon_0V)}$. Consequently, the coupling energy
$\hbar g = |\mathbf{\mathcal{D}} \cdot \mathbf{E}_0|$ between an excitonic dipole $\mathbf{\mathcal{D}}$
and the field $\mathbf{E}_0$ can also be very large. Here, the relevant scale for a ``large'' coupling $g$
is set by the rates at which the energy dissipates from the system, either by spontaneous emission from the
exciton (transverse damping rate $\gamma$), or by leaking out of the cavity (damping rate $\kappa$).
The promised land in cavity quantum electrodynamics (CQED), often called the ``strong-coupling'' regime,
begins, therefore, when $g$ is much larger than both $\gamma$ and $\kappa$.

The incident wave-vector $\mathbf{k}_0$ can be decomposed in perpendicular $\mathbf{k}_z$ and parallel
$\mathbf{k}_{||}$ components in relation to the QW plane ($x-y$ plane). In the growth direction $z$,
the excitons are confined, and therefore, are found in discrete states. In the QW plane, the excitonic
states are found as in bulk, but due to quantum confinement, the degeneracy of the heavy-hole and
the light-hole bands at the center of the band gap is broken. Thus, $\mathbf{k}_0 = \mathbf{k}_{||}
+ \mathbf{k}_z$, where $k_{||}^{2} = k_{x}^{2} + k_{y}^{2} = k_0\sin{\theta}$ ($\theta$ being the
angle between the incident wave-vector and the normal in relation to the surface of the sample)
and $|\mathbf{k}_{z}| = 2\pi n_\mathrm{c}/L_\mathrm{c}$ (for a $\lambda$ cavity). In this case, the cavity dispersion is given by
\begin{equation}
E_\mathrm{c} = \frac{\hbar c}{n_\mathrm{eff}}|\mathbf{k}|= \frac{\hbar c}{n_\mathrm{eff}}\sqrt{\frac{\omega_0^{2}}{c}\sin^{2} {\theta}+ \frac{\pi^{2}}{L_\mathrm{eff}^{2}}}\, ,
\label{Eq:CavityDisp}
\end{equation}
where $n_\mathrm{eff} = \sqrt{n_\mathrm{QW}^{2} -(\kappa/\gamma)}$ is the effective refractive index.

In a semiclassical framework, in order to describe the reflectance spectra of the microcavity,
with an ensemble of $N$ polaritons confined in a SQW, coupled to a single cavity mode, we must
begin considering a driven damped interaction between two coupled oscillators. In this model,
the system is driven by an external radiation field with frequency $\omega$. This field can
interact with complex mixed-mode frequencies close to a resonant mode, in which two states
composed of a cavity-photon (with frequency $\omega_\mathrm{c} - \ri\kappa$) and a 1S
heavy-hole exciton (with frequency $\omega_\mathrm{x} - \ri\gamma$) are coupled with a
matrix element $g$. The damping $\Gamma_0 = (1/4\pi\varepsilon)(\pi/n_\mathrm{c})(e^2/m_0c)f_\mathrm{qw}$
is given by the decay rate of the exciton amplitude at $k=0$ in a SQW, where the exciton has an oscillator
strength $f_\mathrm{qw}$ on the quantum well plane \cite{Andreani}.

Thus, the master equation obtained from the Maxwell-Bloch equation approach, that relates the
phenomenological parameters above to the intra-cavity ($\gamma$) and mirror ($\kappa$) losses
is \cite{Carmichael}
\begin{equation}
[\omega - \omega_\mathrm{x} + \ri(\omega_\mathrm{s} + \gamma)][\omega - \omega_\mathrm{c} + \ri(\omega_\mathrm{s} + \kappa)] = g^2
\label{Eq:Coupledoscill}
\end{equation}
for the eigenvalue frequency $\omega_\mathrm{s}$ of the coupled system. The matrix element $g$ makes
explicit the coupling between the cavity-photon and the exciton, and characterizes the oscillatory
exchange of excitation between the exciton and the cavity field . The mirror loss $\kappa$ (as well as $g$)
is dependent on the reflectivity of the cavity mirrors $R = \sqrt{R_1R_2}$ (composed of two mirrors with
individual reflectance $R_1$ and $R_2$),
\begin{eqnarray}
g &=& \sqrt{\frac{1+\sqrt{R}}{\sqrt{R}}\frac{c\Gamma_0}{n_\mathrm{c} L_\mathrm{eff}}}\,,\\
\kappa &=& \frac{1-\sqrt{R}}{\sqrt{R}}\frac{c}{n_\mathrm{c} L_\mathrm{eff}}\,.
\label{kappa-g}
\end{eqnarray}

The eigenvalues $\omega_\mathrm{s}$ of (\ref{Eq:Coupledoscill}) are given by
\begin{eqnarray}
 \omega_\mathrm{s}^{\pm} &=& \frac{1}{2}\left[\gamma+\kappa +\ri(\omega_\mathrm{x}+\omega_\mathrm{c}-2\omega)\right] \nonumber \\
 &\pm& \frac{1}{2}\sqrt{(\kappa-\gamma)^2 + (\omega_\mathrm{x}-\omega)\Phi_\mathrm{x}+(\omega_\mathrm{c} - \omega)\Phi_\mathrm{c} -4g^2}\,,
\label{Eq:RabiSolution}
\end{eqnarray}
where $\Phi_\mathrm{x} = \omega_\mathrm{x} + 3\omega_\mathrm{c}-4(\omega+\ri\kappa)$ and $\Phi_\mathrm{c}
=\omega_\mathrm{c} + 3\omega_\mathrm{x}-4(\omega+i\gamma)$. These solutions describe the normal modes
formed by the intracavity field and the collective excitonic polarization. In equation (\ref{Eq:RabiSolution})
we identify $\omega_\mathrm{s}^+$ as the excitonic eigenvalue, which expresses the well-known process of
cavity-enhanced spontaneous emission \cite{Purcell}, and $\omega_\mathrm{s}^-$ as the cavity eigenvalue,
which indicates the lesser known companion process of exciton-inhibited cavity decay.

At the limit of the weak intracavity field for a coincident pump frequency $\omega$, cavity resonance
frequency $\omega_\mathrm{c}$ and excitonic transition frequency $\omega_\mathrm{x}$, the eigenvalues
obtained for the coupled system are as follows:
\begin{equation}
\omega_\mathrm{s}^{\pm} = -\frac{1}{2} \left[ (\gamma+\kappa) \pm \sqrt{(\gamma-\kappa)^2 - 4g^2}\right].
\label{Eq:RabiSolResonant}
\end{equation}

When $g$ is small compared to both $\gamma$ and $\kappa$, precisely $2g < |\gamma-\kappa|$, the
square root is purely real. This corresponds to the so-called ``weak-coupling'' regime, which is
usually not considered as CQED. However, when $\gamma \ll g^2/\kappa \ll \kappa$, where $\mathrm{Im}[\omega_\mathrm{s}^\pm]$
is still zero, this is often called the ``bad-cavity'' limit of CQED. At this limit, the cavity decay rate
is the dominant one, but $g^2/\kappa$, which is the excitonic damping induced by exciton-photon coupling,
is larger than the excitonic damping $\gamma$ itself. This means that excitons are more likely to decay
in the cavity mode, rather than in another mode outside the cavity. Since the light can be coupled with
excitons outside the cavity with great efficiency, this regime is interesting for applications where the
field must be extracted from the cavity with great efficiency.

However, when $\mathrm{Im}[\omega_\mathrm{s}^\pm] \neq 0$, precisely $2g > |\gamma-\kappa|$,
one reaches the ``strong-coupling limit'' domain, where $g \gg (\gamma, \kappa)$, and $\omega_\mathrm{s}$
exhibit a normal-mode splitting. In the time domain, this limit corresponds to the appearance of a coherent
exchange of photons, at a rate $g$, between the cavity field and excitons (Rabi-oscillations).
In this context, the Rabi-splitting $\Omega = \omega_\mathrm{s}^+ - \omega_\mathrm{s}^- = 2|\mathrm{Im}[\omega_\mathrm{s}]|$ is given by
\begin{equation}
\Omega = \sqrt{4|g|^2 - (\gamma-\kappa)^2}\,.
\label{Eq:RabiFrequency}
\end{equation}
In this case, we can verify that $\Omega$ may be less than $2g$ due to the broadening.

\begin{figure}[!t]
\centerline{
\includegraphics[width=0.5\textwidth]{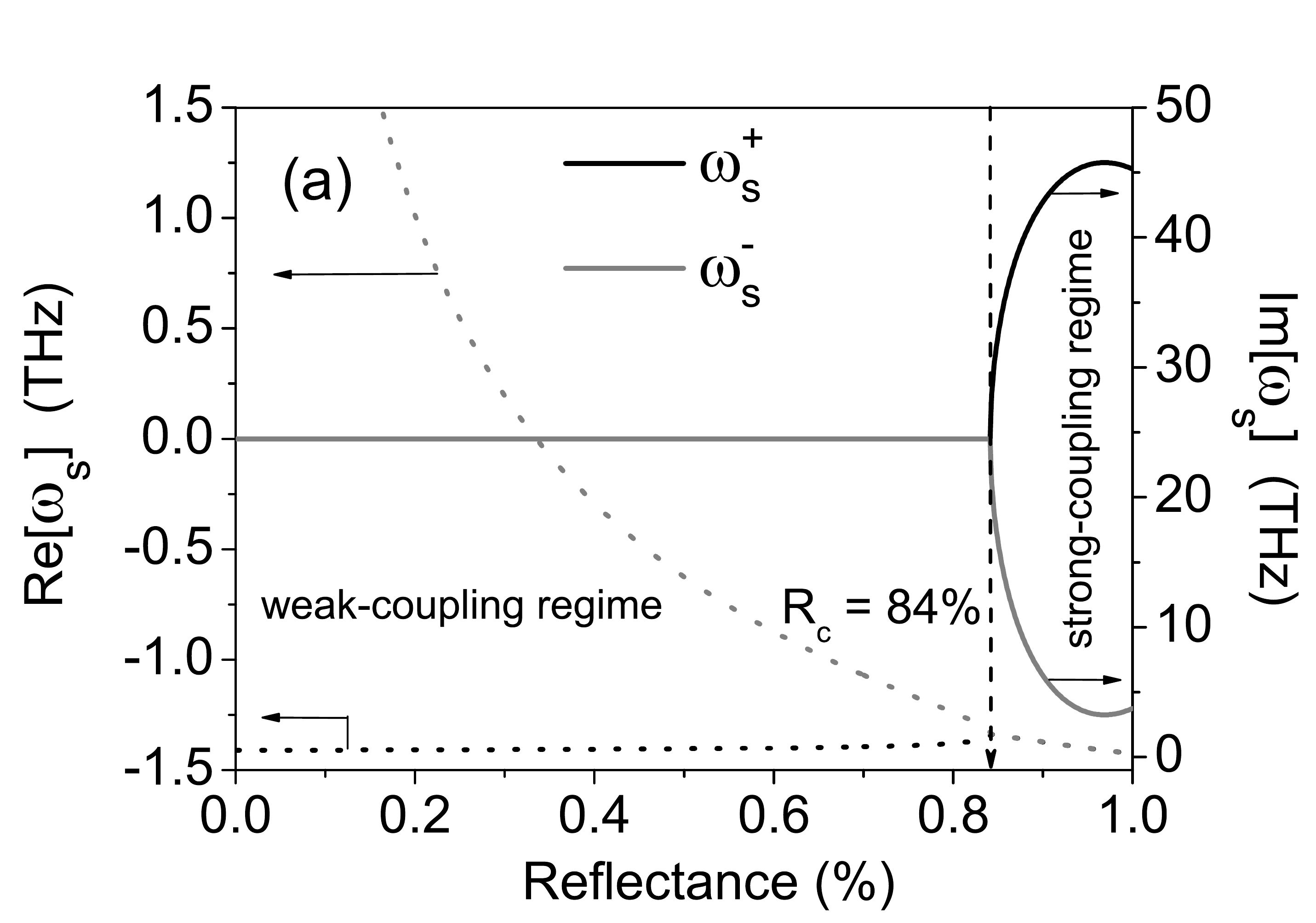} \hfill 
\includegraphics[width=0.48\textwidth]{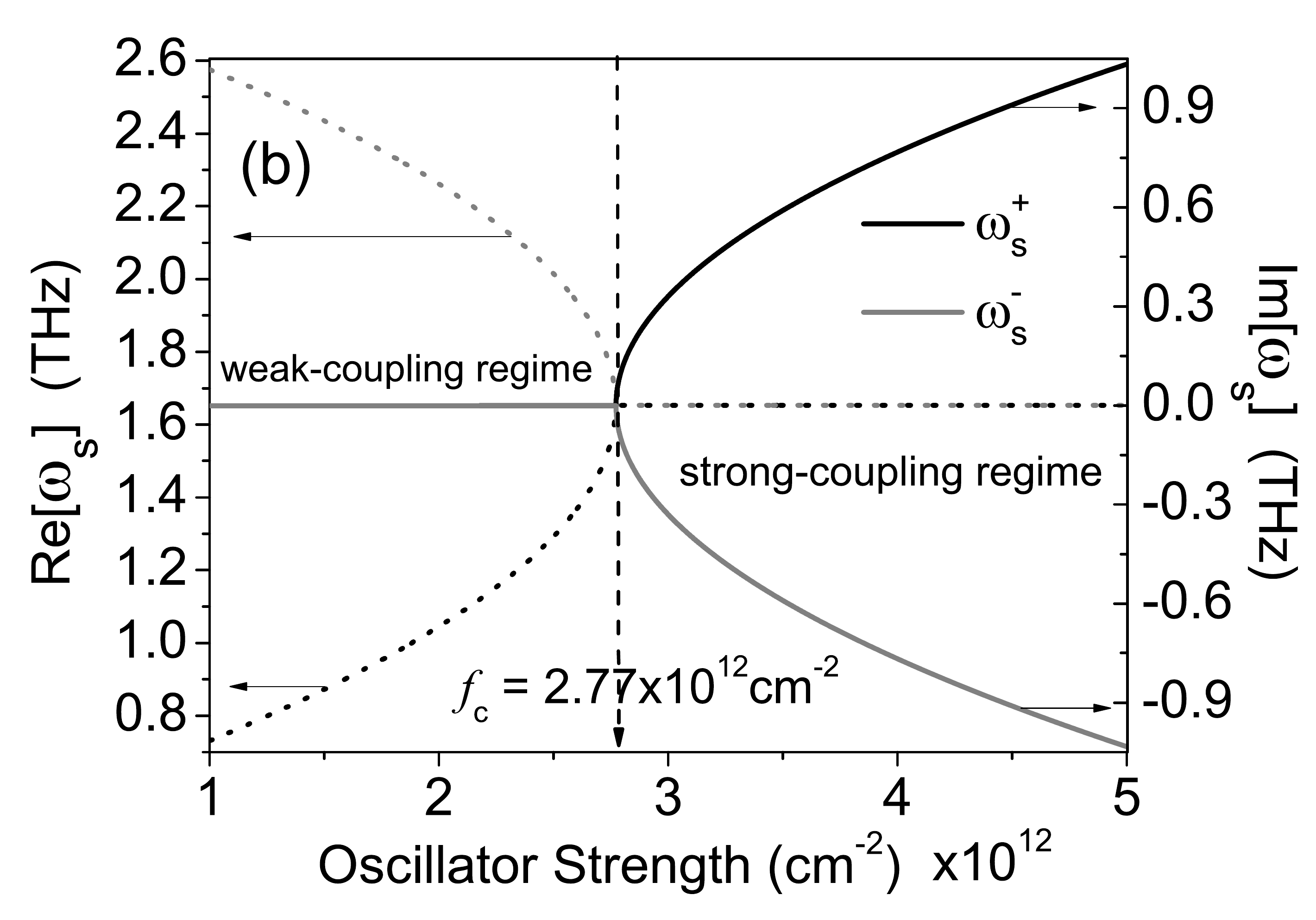}
}
  \caption{Behavior of the real (dotted lines) and imaginary (solid lines) parts of
  $\omega_\mathrm{s}^{\pm}$ (grey line to $\omega_\mathrm{s}^-$ and black line to $\omega_\mathrm{s}^+$)
  for the structure presented in figure~\ref{Fig:sample}~(a). The vertical dashed line limits the strong
  and weak coupling regimes, where the critical reflectance and oscillator strength are
  $R_\mathrm{c} \approx 84\%$ [in figure (a)] and $f_\mathrm{c} \approx 2.77 \times 10^{12}$~cm$^{-2}$
  [in figure~(b)], respectively. The parameters set to produce this result are based on a 100~{\AA}
  GaAs SQW, where $\varepsilon = 12.9$, $\gamma = 0.5$~THz (obtained by the linewidth of the PL spectrum).
  The DBR mirror properties are: $n_H = 3.527$ and $n_L = 3.002$; and $n_\mathrm{c} = 3.458$ (for $\lambda_0 = 800$~nm).}
  \label{Fig:ws}
\end{figure}

\begin{figure}[!b]
\centerline{
\includegraphics[width=0.5\textwidth]{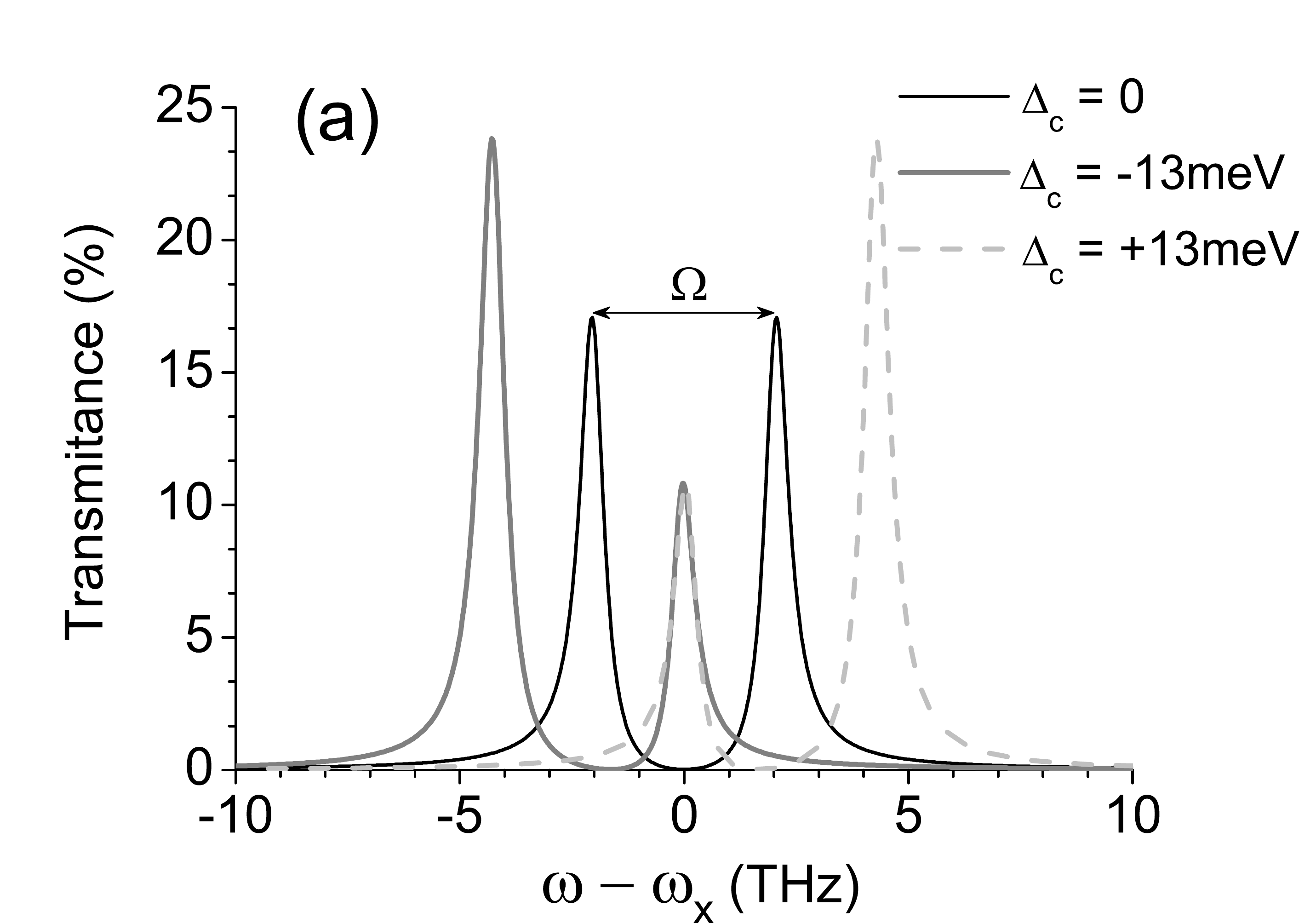}
\hfill
\includegraphics[width=0.48\textwidth]{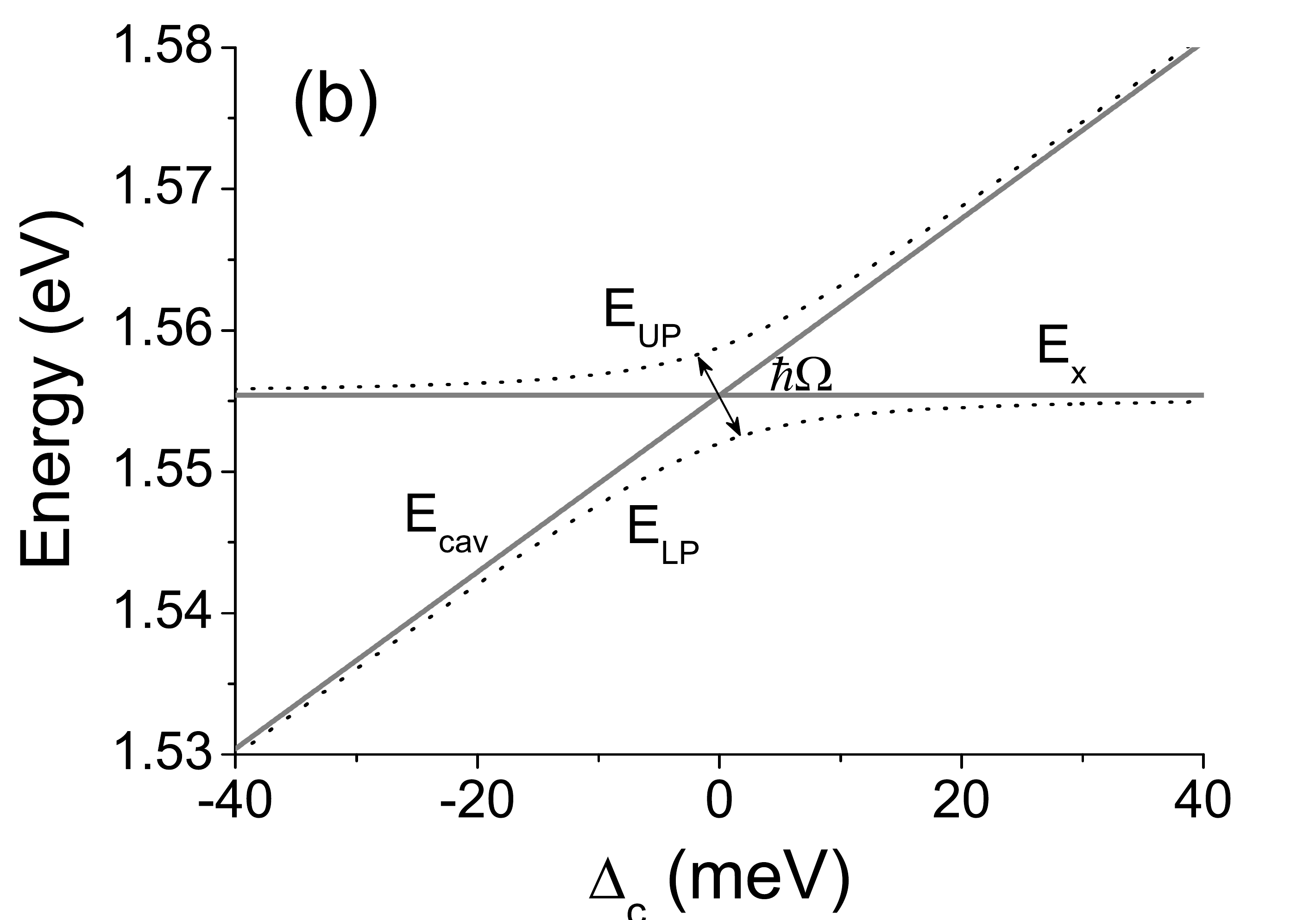}
}
  \caption{(a) Theoretical results for transmission spectra using equations (\ref{Eq:RabiSolution})
  and (\ref{Eq:Transmission}) for three different conditions for cavity detuning.
  The minimum separation between the two peaks occur in $\Delta_\mathrm{c} =0$ and
  (b) The dotted lines show the anti-crossing between upper ($E_\mathrm{UP} = \hbar \omega_\mathrm{s}^+$)
  and lower ($E_\mathrm{LP} = \hbar \omega_\mathrm{s}^-$) polariton branches. The solid grey lines present
  the cavity-photon ($E_\mathrm{c}$) and exciton ($E_\mathrm{x}$) energies.}
  \label{Fig:anticrossing}
\end{figure}

The real and imaginary parts of $\omega_\mathrm{s}^{\pm}$ can be analysed in figure~\ref{Fig:ws}
on the resonance ($\omega = \omega_\mathrm{c} = \omega_\mathrm{x}$). The splitting in $\omega_\mathrm{s}^{\pm}$
occurs in the real part when the system presents it in the weak-coupling regime, and in the imaginary part
when it is in the strong-coupling regime, in which both widths coincide, and are given by $(\kappa + \gamma)/2$.
The crossover occurs approximately at the point where the enhanced exciton decay rate crosses the free
cavity-mode decay rate. The critical reflectivity is approximately given by
$R_\mathrm{c} = 1 - 4\sqrt{2(n_\mathrm{c}L_\mathrm{eff}\Gamma_0/c)}$ and is
found to be $R \approx 84\%$ with the parameters presented in figure~\ref{Fig:ws}~(a).
In figure~\ref{Fig:ws}~(b), we found that a material with an oscillator strength higher than
$f_\mathrm{c} = 2.77\times 10^{12}$~cm$^{-2}$ is necessary to construct a semiconductor microcavity
that presents a strong-coupling regime. In this case, our microcavity is capable of producing Rabi-splitting,
since we are using a SQW of GaAs \cite{Houdre}, where the $f_\mathrm{qw} \sim 4.8\times 10^{12}$~cm$^{-2}$.

The transmittance spectrum for the steady-state coupled system is related to solutions for $\omega_\mathrm{s}$ by \cite{Kimble}:
\begin{equation}
T(\omega)=\left|\frac{\kappa[\gamma + \ri(\omega_\mathrm{x} + \omega_\mathrm{c} - 2\omega)]}{(\omega_{s}^+
+ \ri\delta_\mathrm{c})(\omega_{s}^- + \ri\delta_\mathrm{c})}\right|^2\,,
\label{Eq:Transmission}
\end{equation}
here, $\delta_\mathrm{c} = (\omega_\mathrm{c} - \omega)$ are the detuning between the cavity mode and incident frequency.
The transmission of the cavity shown in figure~\ref{Fig:anticrossing}~(a) is similar to the empty cavity
transmission, except for a double peak on the cavity resonance, separated by $\Omega$. This ``Rabi-splitting''
was observed experimentally in the semiconductor microcavity, as will be discussed in the next section.
In figure~\ref{Fig:anticrossing}~(b) we can observe an anti-crossing between the two peaks of the transmission spectra.


\section{Results and analysis}
\begin{figure}[!t]
  \begin{center}
\includegraphics[width=0.55\textwidth]{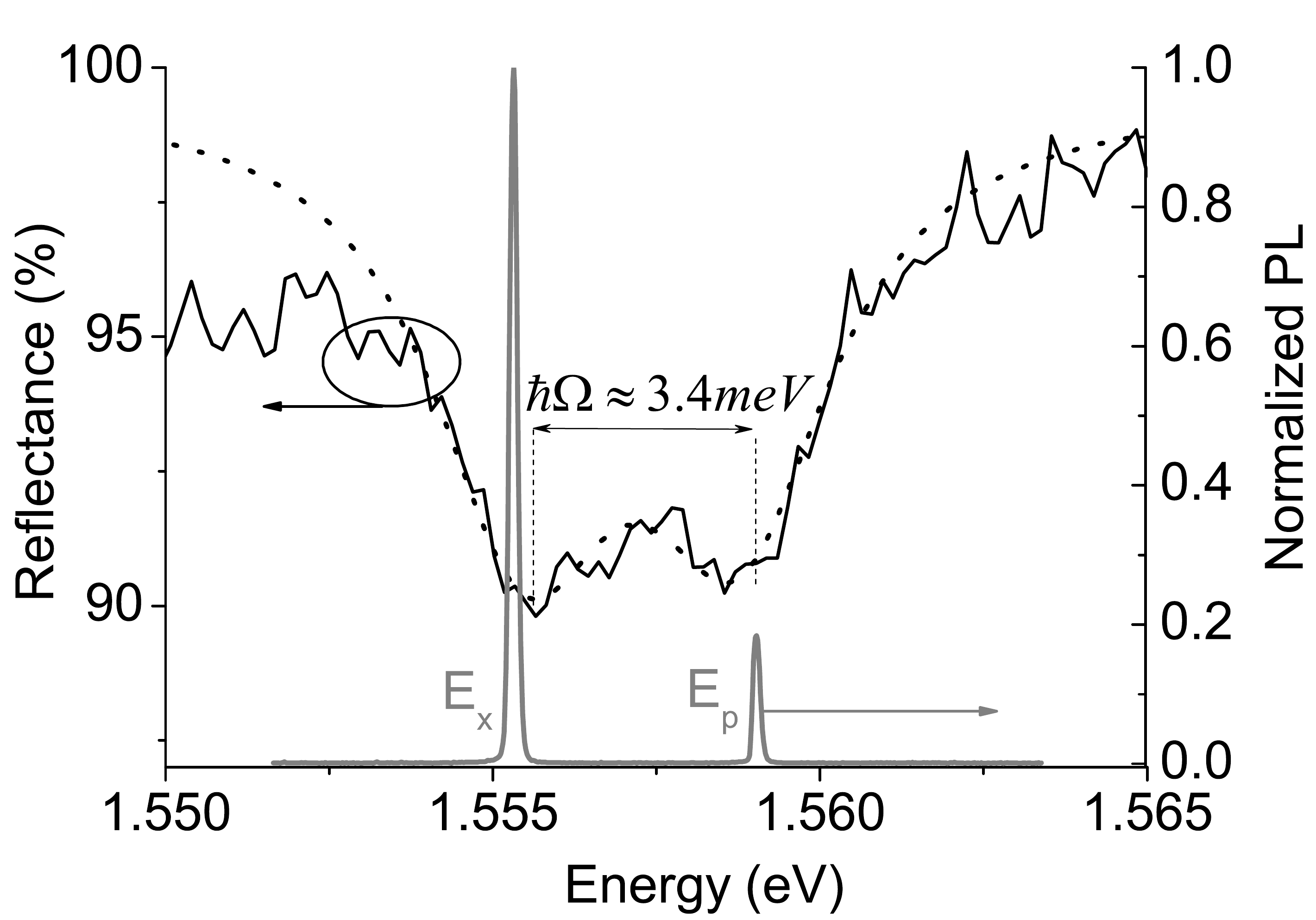}
   \end{center}
  \caption{White light reflectance spectrum at normal incidence around the cavity resonance. The black line is the experimental measurement and the dotted line is the theoretical fitting using the equation (\ref{Eq:Transmission}). The parameters used in the fitting are: ($\kappa ; \gamma$) = ($5.0 ; 5.0$)~THz, $f_\mathrm{qw} = 6.1 \times 10^{12}$~cm$^{-2}$ and $E_\mathrm{c} = 1.5551$~eV. The grey line is the normalized photoluminescence (PL) spectrum of the microcavity, showing the laser emission regime when excited quasi-resonantly by a Ti:Sapphire laser ($E_\mathrm{p}$).}
  \label{Fig:RabiFitt}
\end{figure}

Using the unpolarized white light source we make measurements of reflectivity at 10~K,
in resonant condition, that shows regions in the sample where the strong-coupling regime appears
(see figure~\ref{Fig:RabiFitt}). This regime can be distinguished from the typical cavity resonance
by a double dip, ensuring the exciton-polariton formation. Therefore, a direct measurement of the
spectra gives us a Rabi-splitting energy of $\hbar\Omega = 3.4$~meV. A good fitting agreement in
this figure was reached using the semiclassical coupled oscillator approach, and transfer matrix
method, to determine $\kappa$ and $f_\mathrm{qw}$. The ``bare'' exciton energy $E_\mathrm{x}$ was
fixed in all calculations, and was obtained experimentally by a PL spectrum of the sample using a Ti:Sapphire pumping laser,
in which we found $E_\mathrm{x} = 1.5554$~eV (see the grey line in figure~\ref{Fig:RabiFitt}).
Both lines in the reflectance spectrum are well fitted by Lorentzian lines having the same FWHM
linewidth (full width at half maximum) of $\Delta E \approx 3.5$~meV, indicating a dominating
homogeneous broadening. The linewidth of the UPB corresponds to the nonradiative broadening of
the heavy-hole excitons and the homogeneous broadening mechanisms for the polaritons, e.g.,
their radiative decay, and exciton dephasing due to photon scattering and free-carrier scattering.
The cavity linewidth becomes smaller under resonant conditions for high fineness cavities, because
the coupled-mode linewidth can be smaller than the natural excitonic linewidth \cite{Thompson,Yamamoto} due to QW interaction.

In the strong-coupling regime, the dependence of the coupling factor $2g \approx \Omega$ on
the QW oscillator strength $f_\mathrm{qw}$ (through exciton radiative rate $\Gamma_0$),
permits the use of the fitting procedure, presented in figure \ref{Fig:RabiFitt}, as the
most accurate and reliable measurements of $f_\mathrm{qw}$. In our sample, the value
obtained from the fitting to $f_\mathrm{qw}$ is $6.1\times 10^{12}$~cm$^{-2}$, which is
in good agreement with the literature \cite{Baoping,Andreani-b} taking into consideration an increase of
$f_\mathrm{qw}$ as the temperature decreases and the confinement effect.

\begin{figure}[!t]
  \begin{center}
\includegraphics[width=0.525\textwidth]{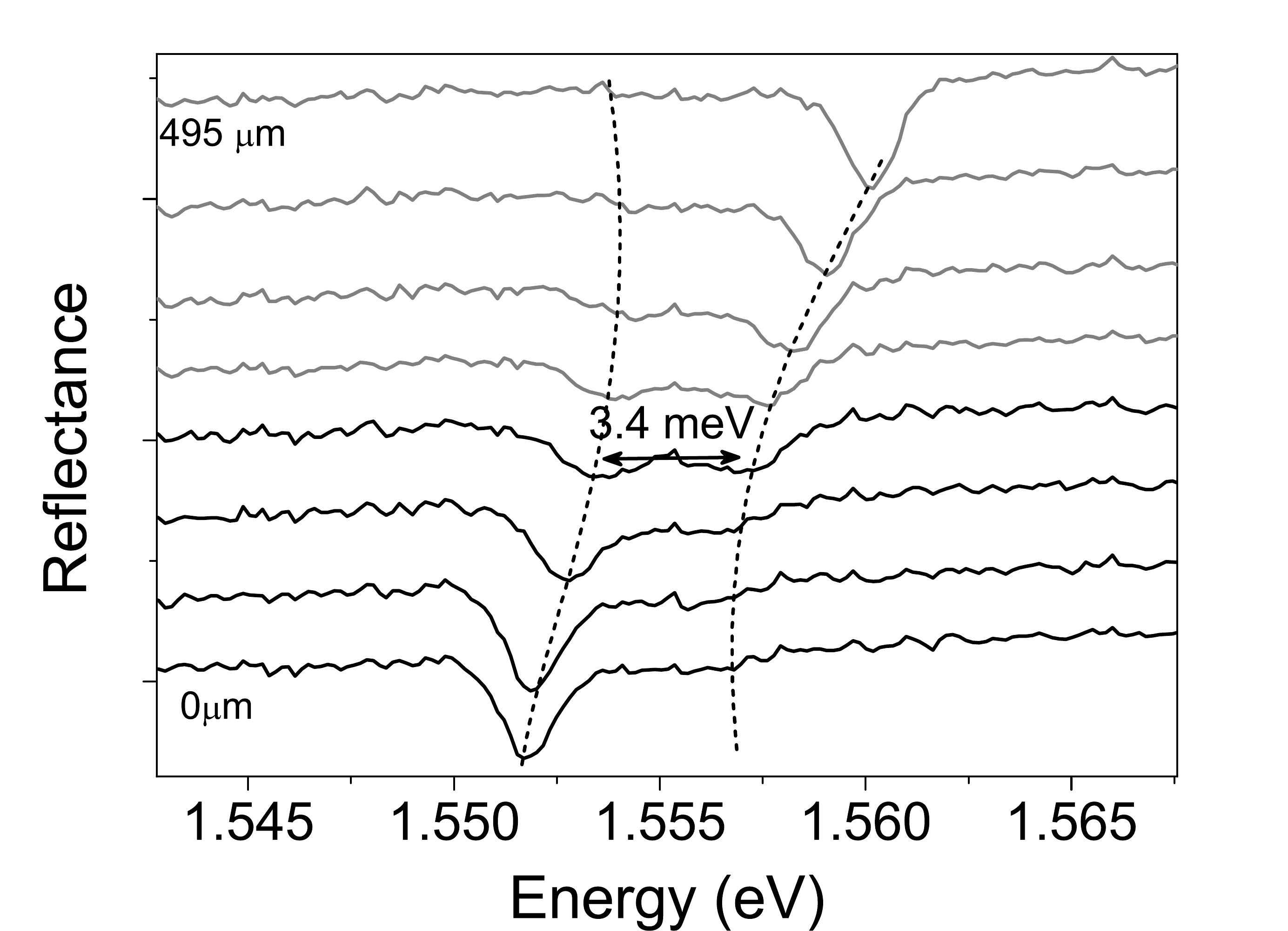}
   \end{center}
  \caption{Formation of cavity polariton mapped by reflectance spectra of the microcavity around the
  cavity resonance for different detunings between exciton and cavity modes. The position of each
  spectrum was changed in equal steps of $90~\mu$m along a line. The black line is for $\Delta_\mathrm{c} \leqslant 0$
  and the grey line for $\Delta_\mathrm{c} > 0$. The dotted line is an eye guide showing the
  anti-crossing theoretically expected in figure~\ref{Fig:anticrossing}~(b).}
  \label{Fig:RabiPos}
\end{figure}

The reflectance spectra for cavity detuning were registered while exciting the sample in different positions on the surface,
as can be viewed in figure~\ref{Fig:RabiPos}. In this figure, we can observe the change from strong to weak
coupling regimes for displacements of about $500~\mu$m along a line with steps of about $90~\mu$m.
Comparing figures~\ref{Fig:CavityDisp}~(a) and \ref{Fig:RabiPos}, we can analyse the behavior of the
Hopfield coefficients $|X_0|^2$ and $|C_0|^2$ (exciton and photon behavior of the polariton),
and identify their weight in the reflectance spectrum in the strong-coupling regime. For $\Delta_\mathrm{c} > 0$
(grey lines) we can verify that the right peak (higher energy) survives, which corresponds to exciton-like behaviour
for the polariton. For $\Delta_\mathrm{c} < 0$ (black lines) the left peak (lower energy) survives, which corresponds
to predominantly photon-like behaviour. If the vacuum-field Rabi-splitting exceeds the original broadening of the
exciton line at the anticrossing condition, the tails of the excitonic distribution weakly couple with the light
remaining between the two split modes and do not effect the reflection.
The dotted line presented in this figure is an eye guide to characterize
the anti-crossing presented in figure~\ref{Fig:anticrossing}~(b); as
theoretically expected, this result provides direct evidence of quantized cavity photon number.

\begin{figure}[!b]
  \begin{center}
\includegraphics[width=0.525\textwidth]{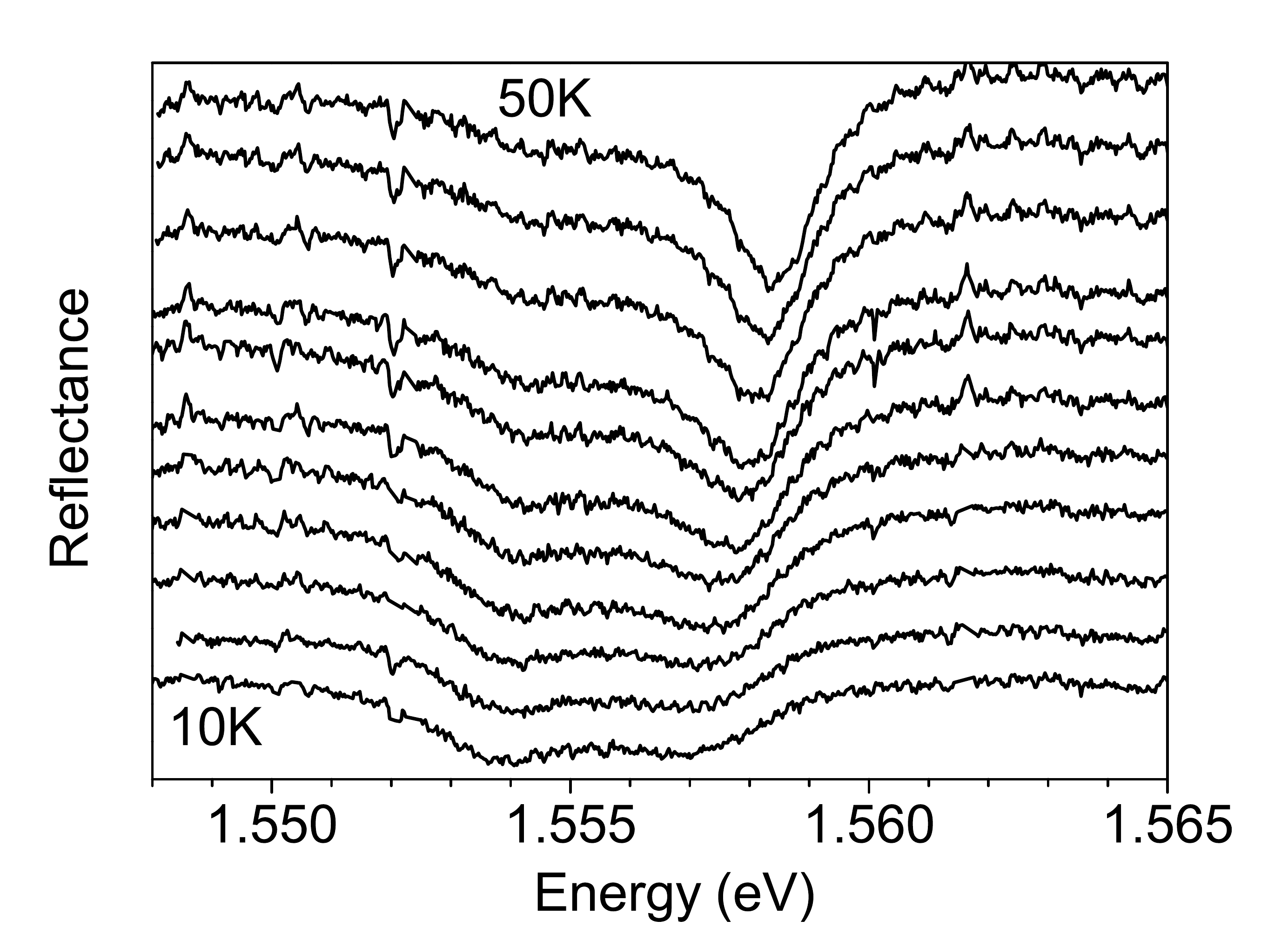}
   \end{center}
  \caption{Reflectance spectra around the cavity resonance at different temperatures in resonant condition.
  The temperature was changed in equal steps from 10~K to 50~K.}
  \label{Fig:RabiTemp}
\end{figure}

We also varied the sample temperature to obtain the temperature-dependent reflectance measurements,
at low white light pump intensities (see figure \ref{Fig:RabiTemp}). In this figure we can verify
the presence of the strong-coupling regime up to the critical temperature of $T_\mathrm{c} \approx 40$~K.
The thermal energy for this temperature ($k_\mathrm{B} T_\mathrm{c}$) is about equal to Rabi energy
($\hbar\Omega$), so that above this temperature the thermal energy is sufficient to break the
exciton-photon coupling, dissociating the polaritons. The exciton binding energy for a SQW 100~{\AA}
GaAs with a barrier of Al$_{0.3}$Ga$_{0.7}$As is about 8~meV \cite{Greene}, which is much higher
than $k_\mathrm{B} T_\mathrm{c}$. Thus, the free-carrier is not broken, but the resonance in the
reflectance spectra that survives for $T > T_\mathrm{c}$ is due to the photon behavior.

All these microcavity concepts and results should influence the conventional design of important
optical and electro-optical devices such as horizontal and vertical-cavity emitting lasers,
electro-optic modulators, and nonlinear optical etalons. In VCSEL's (vertical cavity surface
emitting lasers), a major impact will occur only if laser action is based on exciton recombination.
However, it is well known that, at least at room temperature, exciton dissociation and carrier
interactions are much faster than the exciton radiative lifetime (in $ns$ range), so that they usually
decay in electron-hole pairs. The existence of rapid Rabi oscillation might drastically change this
state of affairs, as the oscillation could be faster than the dissociation time, leading to an efficient
radiative recombination whenever a coupled exciton-photon mode escapes the cavity (polariton lasers).
For nonlinear optical devices which rely on unrelaxed excitation, excitons are still very important
at room temperature and are the root of the unsurpassed performance of QW heterostructure based systems \cite{Weisbuch-b}.
In the near future they should lead to strongly improved device performances.

\section{Conclusions}

In conclusion, we have directly observed exciton-polariton Rabi-splitting of 3.4~meV by reflectance
characterization for only one GaAs single quantum well. A good description of the asymmetries observed
in reflectivity spectra indicating a homogeneous broadening of the free-carrier is described within a
simple semiclassical approach. Compared with the conventional absorption technique, this method allowed
us to study the oscillator strength of excitons confined in a SQW with a higher accuracy. The experimental
results provide a very good agreement with theoretical purpose in which reflectance has been used as an optical probe.

The semiclassical approach allows us to determine the critical reflectance and oscillator strength
necessary to achieve the strong-coupling regime for a general semiconductor microcavity based
on Al$_{x}$Ga$_{1-x}$As technology. Finally, we expect that these experiments will find an interesting
application in new coherent light sources.

\section*{Acknowledgement}
The authors would like to thank F.M.~Matinaga of Federal University of Minas Gerais (UFMG) for the use
of his cryostat and for fruitful discussions of the sample. We would also like to thank the
Instituto Nacional de Ci\^encia e Tecnologia~--- Dispositivos Semicondutores (INCT-DISSE) and
the National Research Council (CNPq) for their financial support.


\ukrainianpart

\title{Визначення сили осциляторів обмежених екситонів \\ у напівпровідниковому
 мікрорезонаторі}

\author{E.A. Котта\refaddr{label1,label2},
П.М.С. Рома\refaddr{label1}}
\addresses{
\addr{label1} Факультет фізики, Федеральний університет штату Амазонас, Манаус, Бразилія
\addr{label2} Національний інститут науки і технології напівпровідникових нанопристроїв,
 Бразилія
}

\makeukrtitle

\begin{abstract}
\tolerance=3000%
Нами досягнуто значного експериментального розщеплення Рабі (3.4 меВ) для
 обмежених поляритонів у плоскому напівпровідниковому $\lambda$ мікрорезонаторі
 для одиночної квантової ями GaAs (10 нм), розміщеної в антивузлі. Явище
 розщеплення Рабi детально обговорюється на основі напівкласичної теорії, коли
 для опису системи використовуються два зв'язані гармонічні осцилятори
 (екситони і фотони). В такий спосіб можна отримати дисперсійну криву
 поляритонів, мінімальне значення для коефіцієнта відбивання
 резонатора і силу осцилятора для досягнення сильнозв'язаного режиму. Цей
 підхід описує ансамбль екситонів обмежених одною квантовою ямою і враховує
 дисипацію. Результати представляють як слабозв'язаний режим з посиленням
 спонтанної емісії, так і сильнозв'язаний режим, коли спостерігається
 розщеплення Рабі на дисперсійній кривій. Теоретичні результати порівнюються з
 експериментальними даними для поведінки коефіцієнта відбивання в резонансних i
 нерезонансних умовах і є дуже точними. Це дозволяє з високою точністю
 визначити силу осциляторів обмежених однією квантовою ямою екситонів.
\keywords мікрорезонатор, розщеплення Рабі, сила осцилятора, сильний
 зв’язок, коефіцієнт відбивання
\end{abstract}
\end{document}